\documentclass[prl,twocolumn,aps]{revtex4}
\usepackage{graphicx}
\usepackage{psfrag}
\usepackage{bm}
\usepackage{epsfig}
\usepackage{amsmath}
\usepackage{amssymb}
\usepackage{psfrag}
\begin{document}
\title{Observing Gibbons-Hawking quantum radiation 
in linearly expanding cigar-shaped Bose-Einstein condensates}
\title{Observing thermal quantum radiation 
in linearly expanding cigar-shaped Bose-Einstein condensates}
\title{Observing quantum radiation from acoustic horizons 
in linearly expanding cigar-shaped Bose-Einstein condensates}
\title{Hawking Radiation from Sonic de Sitter 
Horizons in Expanding \\ Bose-Einstein-Condensed Gases}
\title{Gibbons-Hawking Radiation from Sonic de Sitter 
Horizons in Expanding \\ Bose-Einstein-Condensed Gases}
\title{Gibbons-Hawking Effect and Sonic de Sitter 
Horizons in Expanding \\ Bose-Einstein-Condensed Gases}
\title{Gibbons-Hawking Effect in the Sonic de Sitter 
Space-Time \\ of an Expanding Bose-Einstein-Condensed Gas}
\author{Petr O. Fedichev$^{1,2}$ and Uwe R. Fischer$^1$}
\affiliation{$^{1}$Leopold-Franzens-Universit\"{a}t Innsbruck, 
Institut f\"{u}r Theoretische Physik, Technikerstrasse 25, 
A-6020 Innsbruck, Austria \\
$^2$Russian Research Center Kurchatov Institute, 
Kurchatov Square, 123182 Moscow, Russia}

\begin{abstract} 
We propose an experimental scheme to observe the Gibbons-Hawking effect 
in the acoustic analog
of a  1+1-dimensional de Sitter universe, produced in an 
expanding, 
cigar-shaped Bose-Einstein condensate. 
It is shown that a 
two-level 
system created at the center of the trap,
an atomic quantum dot interacting with phonons, 
observes a thermal Bose distribution at the de Sitter temperature.

\end{abstract}
\maketitle

Cosmology in the early universe 
is a branch of physics which is, for all too obvious reasons, 
removed far from experiment. To use a concise 
formulation of the primary dilemma which cosmology faces, 
there exists nothing like the 
possibility  of ``reproducing'' experiments, 
because there has in fact ``only ever been one experiment, still running, and 
we are latecomers watching from the back'' \cite{pickett}. 
While there can be  no truly experimental cosmology, 
as regards in particular the 
reproduction of the (presumed) extreme conditions
which prevailed in the early stages of the universe, one might look 
for analogous phenomena in condensed matter experiments, 
which {\em are} indeed reproducible, and  
can be done at energy scales smaller by 
many orders of magnitude \cite{Grisha}.

An archetypical model of cosmology is the de Sitter universe, in which 
space is essentially empty, 
and the curvature of space-time is 
due to a nonvanishing cosmological constant $\Lambda$ 
\cite{deSitter}. The de Sitter space-time is used in inflationary models 
of the cosmos \cite{Linde}, displays a cosmological horizon,  
and is associated with a thermal spectrum of particles    
at a temperature $T_{\rm dS} \propto \sqrt{\Lambda}$ \cite{Gibbons}. 

Atomic Bose-Einstein condensates (BECs) \cite{Anglin}
have emerged as one of the most suitable condensed matter
systems for the simulation of quantum phenomena in effective space-times 
\cite{SlowLight,Garay,CSM,Leonhardt,Ourselves,BLV}. 
In the present study, we investigate an effective 1+1--dimensional (1+1D)
de Sitter
universe and the associated thermal phonon spectrum of the 
Gibbons-Hawking type \cite{Gibbons}. 
We examine the propagation of axial low-energy modes in 
expanding, strongly elongated condensates, and find  
that the spectrum consists of one massless (phonon) mode 
and a sequence of massive excitations, moving in an 
effective curved 1+1D space-time. 
It is demonstrated that a
1+1D version of the de Sitter metric \cite{deSitter} can be 
realized by the massless phonon mode 
in a linearly expanding Bose-Einstein condensate with 
constant particle interaction. 
We show that an atomic quantum dot (AQD) \cite{AQD},  
placed at the center of the cloud, can be used to measure the 
Gibbons-Hawking quantum process. The detector's coupling to the superfluid is 
constructed in such a way that the natural time interval of the 
detector is equal to the time interval in the de Sitter metric. Therefore, the
detector inside the expanding condensate is capable of measuring a thermal 
state at the de Sitter temperature.
The Gibbons-Hawking effect is a curved space-time generalization 
of the Unruh-Davies effect, in which a constantly accelerated 
detector in vacuum responds as if it were placed in a thermal 
bath with temperature proportional to its acceleration 
\cite{unruh76,BirrellDavies}. 
Therefore, our proposal 
represents  
a means to confirm the observer dependence of the 
particle content of a quantum field in curved space-time \cite{fulling}.

Our approach is based on the by now 
well established identity of the action of a massless scalar 
field propagating on a curved space-time background in $D+1$ dimensions
and the action of the phase fluctuations $\Phi$  
in a moving inhomogeneous superfluid \cite{unruh,Matt,PGRiemann}
(we set $\hbar = m =1$, where $m$ is the mass of a superfluid particle): 
\begin{eqnarray}
S  & = & \int d^{D+1}x 
\frac{1}{2g} \left[ -\left(\dot \Phi -{\bm v} 
\cdot \nabla \Phi\right)^2 + c^2  (\nabla \Phi)^2 \right] \nonumber\\
& \equiv & \frac12 \int d^{D+1}x 
\sqrt{-{\sf g}} {\sf g}^{\mu\nu} 
\partial_\mu \Phi \partial_\nu \Phi \,. \label{action}
\end{eqnarray}
Here, $\bm v ({\bm x},t) $ is the superfluid background 
velocity, $c({\bm x},t) =\sqrt{g\rho_0}$ the velocity of sound, 
$\rho_0({\bm x},t)  $ the background density, 
and $g=4\pi a_s$ is the coupling constant describing the short-range 
interaction between the particles in the superfluid; $a_s$ is the 
$s$-wave scattering length.
In the second line of (\ref{action}), the conventional 
hydrodynamic action is identified with the action 
of a minimally coupled scalar field in an effective curved space-time.  
In the following, we derive the effective 1+1D action 
of the form (\ref{action}) for the lowest 
axial excitations of an elongated condensate from the full 3D dynamics
of hydrodynamic excitations of the cigar-shaped condensate, and identify the 
metric tensor ${\sf g}_{\mu\nu}$. 

\begin{center}
\begin{figure}[bt]
\psfrag{zH}{$z_{\rm H}$}
\psfrag{-zH}{$-z_{\rm H}$}
\psfrag{b}{\large $b\propto t$}
\psfrag{bperp}{\large $b_\perp\propto \sqrt t$}
\psfrag{Atomic}{Atomic}
\psfrag{Quantum Dot}{Quantum Dot}
\vspace*{1em}
\centerline{\epsfig{file=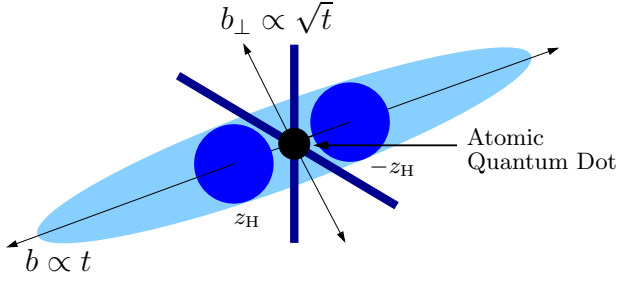,width=0.44\textwidth}}
\caption{\label{Fig1} Expansion of a cigar-shaped Bose-Einstein 
condensate.
The stationary horizon surfaces 
are located at $\pm z_{\rm H}$, respectively. 
The thick dark lines represent lasers creating an optical potential 
well in the center.} 
\end{figure}
\end{center}

We consider a large condensate confined in a strongly 
anisotropic harmonic trap, characterized by the frequencies 
$\omega_\parallel$ and $\omega_{\perp}$ in the axial and radial directions, 
respectively ($\omega_{\perp}\gg \omega_\parallel$). 
The initial condensate density 
is given by the usual Thomas-Fermi (TF) expression  
$
|\Psi_{\rm TF}|^2 
=\rho_m (1-{r^2}/{R_{\perp}^2}-{z^2}/{R_\parallel^2}). 
$ 
Here, $\rho_m$ is the maximum initial density 
and the squared TF radii are $R^2_\parallel= 2\mu/\omega_\parallel^2 $ 
and $R^2_{\perp}=2\mu/\omega_{\perp}^2$, where 
$\mu=\rho_m g$ is the (initial) chemical  potential.
According to \cite{Scaling}, 
the temporal evolution of a Bose-Einstein-condensed atom 
cloud under variation of the trapping frequencies is 
described by the scaling condensate wave function
\begin{equation}
\Psi \left(r/{b_\perp}, z/b \right)= 
\left( 
\rho_0  +\delta\rho \right)^{1/2} 
\exp\left[i (\Phi_0 +\Phi) 
\right] . \label{Wavefunction}
\end{equation}
Here, $\Phi_0 =-\int g \rho_0 ({0},t) dt +{\dot b z^2}/{2b}  
+{\dot b_\perp r^2}/{2b_\perp}$ is  the background phase, where  
$b_{\perp}(t)$ and $b(t)$ are the scaling parameters describing the condensate
expansion in the radial ($\hat r$) and axial ($\hat z$) directions, 
cf. Fig.\,\ref{Fig1}. The mean-field condensate density 
$\rho_0=|\Psi_{\rm TF}|^2 (r/b_\perp,z/b)/(b_\perp^2 b)$ 
contains the {scaling volume} $b_\perp^2 b$, and 
$\delta \rho$ are the fluctuations of the density around mean-field.

\vspace*{-1.5em}
\vspace*{-1.5em}

The elementary excitations above the ground state of the BEC in the limit 
$\omega_{\perp}/\omega_\parallel \rightarrow \infty$ 
were first studied in \cite{Zaremba}.
The description of the modes is based 
on an adiabatic separation for the axial and
longitudinal variables of the phase fluctuation field: 
\begin{equation}
\label{varsep}
\Phi (r,z,t) = \sum_n \phi_n(r) \chi_n(z,t), 
\end{equation}
where $\phi_n(r)$ is the radial wavefunction characterized by the 
quantum number $n$ (we consider only zero angular momentum modes). 
For long wavelength axial  excitations,  
the dispersion relation reads \cite{Zaremba} 
\begin{equation}\label{zarembadisprel}
\epsilon^2_{nk} = 2\omega_{\perp}^2 n (n+1)+c_0^2 k^2, 
\end{equation}
where $c_0 =\sqrt{\mu/2} $ and $k$ is the axial wavenumber 
($kR_\perp \ll 1$).
In the action (\ref{action}), 
we use Eq.\,(\ref{varsep}), with the rescaled radial wavefunction 
$\phi_n\equiv\phi_n(r/b_\perp)$,
and integrate over the radial coordinate. 
We then find the following effective action for the
axial modes of a given radial quantum number $n$, 
\begin{eqnarray}
S_n  &= & \int dt  dz \frac{ b_\perp^2 C_n}{2g}
\left[ -\left(\dot \chi_n -{v}_z \partial_z \chi_n\right)^2 
\right. \nonumber\\& & \qquad  \qquad \quad \qquad\left. 
+ \bar c^2_n (\partial_z \chi_n)^2 + M^2_n \chi_n^2 \right].
\label{1daction}
\end{eqnarray}
Here, the common factor $C_n(z)$ is given by  
$
b_\perp^2 C_n(z)=\int_{r<r_m} d^2r \phi^2_n , 
$ 
%
the averaged speed of sound is 
$
\bar c_n^2(z,t)= {g}{C_n}^{-1}b_\perp^{-2} 
\int_{r<r_m} d^2r \rho_0 
\phi^2_n 
$, 
and the effective mass term
$
M_n^2(z,t)={g}{C_n }^{-1} b_\perp^{-2} 
\int_{r<r_m} d^2r \rho_0 
[\partial_r\phi_n 
]^2$; the radial cigar size is given by 
$r^2_m=R_{\perp}^2b^2_{\perp}(1-z^2/R_\parallel^2b^2)$. 
The phonon branch of the excitations corresponds to the $n=0$ 
solution of Eq.\,(\ref{zarembadisprel}), for which 
the  radial wavefunction $\phi_0=$ const. 
\cite{Zaremba}; hence the mass term $M_0=0$ and 
$
C_0(0)=\pi R_{\perp}^2 
$, 
 $\bar c (0,t) \equiv \bar c_{n=0} (0,t) = c_0 /(b_\perp b^{1/2}).
$ 

The action (\ref{1daction}) can be identified with the action of a minimally
coupled scalar field in 1+1D, close to the center of the condensate. 
The line element 
reads \cite{unruh,Matt}
\begin{equation}
ds^2 = A_c \left[-(c^2 -v_z^2)dt^2 -2v_z dt dz +dz^2 \right], 
\end{equation} 
where $A_c$ is some arbitrary (space and time dependent) 
conformal factor and we set $c=\bar c (z=0,t)$.
The actions (\ref{action}) and (\ref{1daction}) can be made consistent 
if we renormalize the phase field according to $\Phi = Z \tilde \Phi $ 
{\em and}  require that $Z^2 b_\perp^2 C_0 (0)/g =   1/\bar c(0) $, leading to
\begin{equation}
B\equiv 
\frac{b_\perp}{b^{1/2}} =  8 \sqrt{\frac{\pi}2} 
\frac1{Z^2} 
(\rho_m a_s^3)^{1/2} 
\left(\frac{\omega_\perp}\mu\right)^2 = {\rm const}. 
\label{cond1}
\end{equation}
The constant factor $Z$ in the above relations  
does not influence the equation of motion, and
corresponds  to a renormalization of the field amplitude,
$\Phi = Z \tilde \Phi $, which  determines the coupling of 
$\tilde \Phi$ to a detector \cite{No3}.

We require that the space-time metric is, in appropriately chosen
coordinates, identical to that of a 1+1D de Sitter universe.  
We first apply the transformation $c_0 d\tilde t = c(t) dt$,  
connecting the laboratory time $t$ to the time variable $\tilde t$.
Defining $v_z/c = \sqrt \Lambda z = (B \dot b/c_0) z$ 
(note that the ``dot'' on $b$ and other quantities always refers
to ordinary laboratory time),
this results, up to the conformal factor $A_c$, in the line element  
$d s^2 =-c_0^2(1- \Lambda z^2)d \tilde t^2 -2c_0 z \sqrt \Lambda d\tilde t 
dz+dz^2$.
We then employ a second transformation 
$c_0 d\tau= c_0 {d\tilde t} +z\sqrt \Lambda dz/(1-\Lambda z^2)$, with a
time independent $\Lambda$, 
to obtain the 1+1D de Sitter metric
\cite{deSitter,Gibbons}
\begin{equation}
d s^2  = -c_0^2 \left(1-\Lambda  z^2 \right)  
d \tau^2 + \left({1-\Lambda z^2 }\right)^{-1} 
d z^2\,.
\label{deSitterlineelement}
\end{equation}
The quantity $\Lambda = B^2\dot b^2/c_0^2$ is a ``cosmological  
constant,'' provided $\dot b$ does not depend on $t$. 
The transformation between $t$ and the de Sitter time $\tau$, 
at a given coordinate $z$, then involves the exponential ``acceleration,'' 
$
{t}/{t_0} = \exp[B \dot b \tau], 
$ 
where $t_0 \sim 2\pi /\omega_\parallel $ 
is set by the initial conditions. 
The (conformally invariant) temperature associated with the effective 
metric (\ref{deSitterlineelement}) is the Gibbons-Hawking 
temperature \cite{Gibbons} 
\begin{eqnarray}
T_{\rm dS} 
& = &  \frac{c_0}{2\pi } \sqrt{\Lambda} = \frac B{2\pi}  {\dot b}  
\, ,\label{TdS}
\end{eqnarray}
and the 
horizon(s) are located at 
$
z= \pm z_{\rm H}= \pm {R_\parallel} ({\omega^2_\parallel}/
{2\mu\Lambda})^{1/2} . 
$
Combining the latter relation with (\ref{TdS}), 
we see that $z_{\rm H}/R_\parallel \ll 1$, 
if $T_{\rm dS} \gg \omega_\parallel /4\pi $. 
We are thus justified 
in neglecting the $z$ dependence in $C_0$ and $\bar c$ to  describe   
the physics inside the horizon surfaces. 

According to Eq.\,(\ref{1daction}), the equation of motion 
$\delta S_0 /\delta \chi_0=0$  is given, at constant $B$, by
\begin{equation}
B^2 b^2 \frac d{dt} \left( b^2 \frac d{dt}\chi_0 \right) 
- \frac1{C_0 (z_b)}\partial_{z_b} \left[{\bar c^2(z_b) C_0 (z_b)}
\partial_{z_b} \chi_0\right]  \label{chiEqmotion}
= 0,
\end{equation}
where $z_b = z/b$ is the scaling coordinate. Apart from the  
factor $C_0(z_b)$, stemming from averaging over the perpendicular direction,
this equation corresponds to the hydrodynamic equation of phase
fluctuations in inhomogeneous superfluids \cite{Stringari}.

At $t\rightarrow -\infty$, the condensate is in equilibrium and the
quantum vacuum phase fluctuations close to the center of the condensate can 
be written in the following form 
\begin{equation}
\hat\Phi \equiv {\hat \chi_0} 
= \sum_k \sqrt{\frac{g}{4C_0(0) R_\parallel \epsilon_{0k}}}
\,\hat a_{k} e^{ 
-i\epsilon_{0k} t + ikz }
+ {\rm H.c.} ,
\end{equation}
where $\hat a_k,\hat a^\dagger_k$ are the annihilation and creation 
operators of a phonon. The intial quantum state of phonons, which we call the
``adiabatic vacuum,'' is the ground state 
of the superfluid and is annihilated by the operators 
$\hat a_k$ (this definition of vacuum and excitations  
corresponds to the ``adiabatic basis'' of \cite{BirrellDavies}). 
Assuming that the expansion is switched on adiabatically,  
the density fluctuation operator, which is conjugate to 
$\hat\chi_0$, then reads
\begin{eqnarray}
\delta\hat\rho & = & \sum_k
\sqrt{\frac{\epsilon_{0k}}{4C_0(0) R_\parallel g}} 
\nonumber\\& & 
\times 
\frac  d{dt}  \left\{ 
\hat a_{k} \exp \! \left[
-i\int^t\! \frac{dt' \epsilon_{0k}}{Bb^2}+ ik z_b \right]\right\} \!
+ {\rm H.c.} \label{deltarho} 
\end{eqnarray}
The above equation 
completely characterizes 
the $n=0$ evolution of the condensate density fluctuations.

\begin{center}
\begin{figure}[t]
\psfrag{U}{\large $U$}
\psfrag{Omega}{\large $\Omega$}
\psfrag{Vopt}{\large $V_{\rm opt}$}
\psfrag{l}{\large $l$}
\psfrag{alpha}{\large $|\alpha \rangle$}
\psfrag{beta}{\large $|\beta \rangle$}
\psfrag{Delta}{\large $\Delta$}
\vspace*{-8em}
\centerline{\epsfig{file=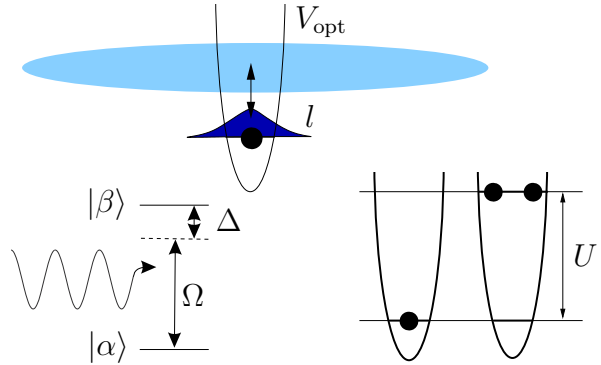,width=0.44\textwidth}}
\caption{\label{Fig2} 
Level scheme of the ``Atomic Quantum Dot,'' which is embedded in 
the superfluid cigar, and created by an optical well for atoms
of a hyperfine species different from that of the condensate. 
Double occupation of the dot is prevented by a 
collisional blockade mechanism.}
\end{figure}
\end{center}

The particle content of a quantum field state in curved space-time
depends on the observer \cite{fulling}. 
We now show that the de Sitter time interval 
$d\tau=dt/bB=dt/b^{1/2}b_\perp$ 
can be effectively measured by an {\em Atomic Quantum Dot} \cite{AQD}.
Here, the AQD is an effective
two-level system with a time-dependent level splitting (see Fig.\,\ref{Fig2}).
It can be made in a gas of atoms possessing  two hyperfine
ground states $\alpha$ and $\beta$. The atoms in the state $\alpha$ 
represent the superfluid cigar,
and are used to model the expanding universe. 
The AQD itself is formed by trapping atoms in the state $\beta$ in a 
tightly confining optical potential
$V_{\rm opt}$. The interaction of atoms in the two internal levels is
described by a set of coupling parameters $g_{cd} = 
4\pi a_{cd}$ ($c,d =\{\alpha,\beta\}$),
where $a_{cd}$ are the $s$-wave scattering
lengths characterizing short-range intra- and inter-species collisions;
$g_{\alpha\alpha}\equiv g$, $a_{\alpha\alpha} \equiv a_s$.
The on-site repulsion between the atoms $\beta$ 
in the dot is $U\sim g_{\beta \beta}/l^{3}$,
where $l$ is the characteristic size of the ground state wavefunction
of atoms $\beta$ localized in $V_{\rm opt}$. In the following we consider the collisional
blockade limit of large $U>0$, where only one atom of type
$\beta$ can be trapped in the dot. This assumes that $U$ is larger
than all other relevant frequency scales in the dynamics of both
the AQD and the expanding superfluid. (The value of  
$g_{\beta\beta}$ may be increased using Feshbach resonances \cite{feshbach}.)
As a result, the appropriate collective coordinate of the AQD 
is modeled by a 
pseudo-spin-$1/2$, with 
spin-up/spin-down state corresponding to occupation by a single/no atom
in state $\beta$.

\vspace*{-2em}

Conversion of atoms between the 
states $\alpha$ and $\beta$ is caused by a 
laser-driven transition characterized by Rabi frequency
$\Omega$ and detuning $\Delta$. The Hamiltonian of the AQD interacting
with the superfluid is 
\begin{eqnarray}
H_{\rm AQD} & = & 
\left\{-\Delta+g_{\alpha\beta}[\rho_0(0,t)+\delta\rho(0,t)]\right\}
\frac{\left(1+\sigma_{z}\right)}2\label{HAQD}\\
&  & +\Omega \left(\rho_0(0,t) l^{3}\right)^{1/2} 
\exp[i(\Phi+\Phi_0)]\sigma_{+}+ {\rm H.c.},  \nonumber
\end{eqnarray}
where we use Pauli matrix notation, 
i.e. $\sigma_+ = \sigma_x + i\sigma_y$.
We introduce the wavefunction of the AQD in the form 
$\psi=\psi_\beta\exp[i(\Phi+\Phi_0)]|\beta\rangle+\psi_\alpha|\alpha\rangle$.
Using the scaling expression (\ref{Wavefunction}) for the condensate
wave function, we find
\begin{eqnarray}
i\frac{d\psi_\beta}{d\tau} & = & \left\{bB[-\Delta
+\rho_0 (g_{\alpha\beta}-g)]+\delta V \right\}\psi_\beta 
+\frac{\omega_{0}}2 \psi_\alpha, \nonumber\\
i\frac{d\psi_\alpha}{d\tau} & = &  \frac{\omega_{0}}2 \psi_\beta,
\end{eqnarray}
where $\tau$ is the \emph{de Sitter time}, and 
$\omega_{0}=2\Omega(\rho_m l^{3})^{1/2}$ is independent of $\tau$.
The perturbation potential 
\begin{equation}
\delta V (\tau) = (g_{\alpha\beta} -g)  B b(\tau)
\delta\rho({0},\tau) \label{deltaV}
\end{equation} 
provides the coupling between the AQD and the expanding superfluid
(if $g_{\alpha\beta}\simeq g$, higher order terms in the density 
fluctuations have to be taken into account 
in the Rabi term of (\ref{HAQD})).  
We suggest to operate the detector 
at $\Delta (t) =(g_{\alpha\beta}-g)\rho_{0}(0,t)=
(g_{\alpha\beta}-g)\rho_m/(\dot b^2 B^2 t^2)$,
so that in zeroth order in $\delta V$ we obtain an effective two-level
system with the level splitting $\omega_{0}$, which 
plays the role of a frequency standard of the
detector. 
By adjusting the laser intensity contained in 
the Rabi frequency $\Omega$, one can change 
$\omega_0$, and thus probe the response of the detector for 
various frequencies. 

The detector response can be calculated using the two-point correlation
function of the coupling operator $\delta \hat V$. The probabilities for 
the detector to ``click,'' corresponding to 
excitation ($P_+$) and de-excitation ($P_-$) of the detector, 
per unit of de Sitter time, are given by \cite{unruh76,BirrellDavies}:
\begin{equation}
\frac{dP_{\pm}}{d\tau}=\lim_{{\cal T}\rightarrow\infty}
\frac{1}{\cal T}\int^{\cal T}\!\!\!\int^{\cal T} 
\! d\tau d\tau^{\prime}\langle\delta \hat V(\tau)\delta \hat V(\tau^{\prime})
\rangle 
e^{\mp i\omega_{0}(\tau-\tau^{\prime})}.
\label{detectorresponse}
\end{equation}
Using the explicit form of the detector-to-field coupling (\ref{deltaV}),
we obtain 
\begin{equation}
\langle\delta \hat V(\tau)\delta \hat V(\tau^{\prime})\rangle
\propto { (\rho_m a_s^3)^{1/2}} \left(\frac{\omega_\perp}\mu\right)^2
\frac{B^2(g_{\alpha\beta}/g-1)^2 T_{\rm dS}^{2}}{\sinh^{2}
[\pi T_{\rm dS}(\tau-\tau^{\prime})]}.\label{correlator}
\end{equation}
This 
correlation function is proportional to one
characterizing a thermal phonon state 
at a temperature $T_{\rm dS}$ in a condensate at rest. This means that our 
de Sitter detector
at the center of the condensate responds to the scaling vacuum 
as if it were placed in a thermal bath at a temperature $T_{\rm dS}$. 
Indeed, substituting the
correlator (\ref{correlator}) into Eq.\,(\ref{detectorresponse}) we
find that the quantities $dP_{\pm}/d\tau$ at late times 
are time independent, and satisfy detailed balance conditions 
corresponding to thermal
equilibrium at the de Sitter temperature given in Eq.\,(\ref{TdS}):
\begin{equation}
\frac{dP_{+}/d\tau}{dP_{-}/d\tau} = \frac{n_{\rm B}}{1+n_{\rm B}}, 
\label{occup}
\end{equation} 
where $n_{\rm B}=(\exp[\omega_{0}/T_{\rm dS}]-1)^{-1}$ 
is the Bose distribution function.
We stress that the AQD observes a thermal spectrum, even though 
the quantum state of the fluctuations is adiabatically connected 
to the initial vacuum state, i.e., no excitations in the adiabatic 
basis characterized by the $\hat a_k$ are created 
(note that the time interval of the adiabatic vacuum 
is $dt / B b^2 = d\tau/b$).
The latter process 
corresponds to ``cosmological'' quasiparticle production 
\cite{Ourselves,BLV}, which 
can be neglected in our situation.

Our experimental proposal consists in the following steps: 
(a) Preparation of a large TF condensate at very low temperatures. 
(b) Introduction of an AQD at the center (respectively an 
array of AQDs to increase the signal), which can be used
to determine the initial temperature of the cloud.
(c) Linear axial expansion of the condensate according to (\ref{cond1}).
The Rabi oscillations of the AQD can be used to monitor the
thermal distribution $n_{\rm B}$ in (\ref{occup}). 
Since $Z^2\propto (\rho_m a_s^3)^{1/2} 
\left({\omega_\perp}/\mu\right)^2$ and $P_\pm \propto Z^2$ \cite{No3}, 
the initial condensate has to be quite 
dense and close to the quasi-1D r\'egime.
The particle density throughout the cloud decreases like $t^{-2}$. 
Therefore, the rate of three-body losses 
quickly decreases during expansion, and 
comparatively long observation times are feasible. 

The AQD as a detector 
is suitable to measure the de Sitter time interval,
because it couples linearly to the square root of the  
density of the superfluid gas. In principle, one can construct detectors
coupling to different powers of density or superfluid velocity, and 
then more generally experimentally 
study the non-uniqueness of the particle content of various
quantum states in curved space-time \cite{fulling}.
For example, 
outcoupling pairs of atoms by photoassociation 
is a means to set up a detector which has $d\tau/dt \propto \rho_0$.  
Finally, we note that while in this paper we have studied only the 
$n=0$ massless axial phonon modes,  
strongly elongated condensates can also be used 
to study the evolution of massive bosonic excitations.
Together with a natural Planck scale 
$E_{\rm Planck} \sim \mu$, this provides the opportunity 
to investigate, on a laboratory scale, 
the influence of finite quasiparticle mass 
and the trans-Planckian spectrum on the 
propagation of relativistic quantum fields in curved space-time.

We thank E.\,A. Cornell for an inspiring discussion on the 
experimental feasibility of our theoretical ideas. 
We acknowledge helpful discussions with 
R. Parentani, R. Sch\"utzhold,  G.\,E. Volovik, 
P. Zoller,  and A. Recati. 
P.\,O.\,F. has been supported by the Austrian 
FWF and the 
Russian 
RFRR, 
and U.\,R.\,F. by the FWF.
We gratefully acknowledge support from the ESF Programme 
``Cosmology in the Laboratory.''


\begin{thebibliography}{499}
\bibitem{pickett} G. R. Pickett {\it et al.}, 
Nature {\bf 383}, 570 (1996).
\bibitem{Grisha} G.\,E. Volovik, 
{\it The Universe in a Helium Droplet} (Oxford University Press, Oxford, 
2003).
\bibitem{deSitter} W. de Sitter,  
Mon. Not. R. Astron. Soc. {\bf 78}, 3  (1917).
\bibitem{Linde} A. Linde, 
{\em Inflation, Quantum Cosmology and the Anthropic Principle}, 
hep-th/0211048.
\bibitem{Gibbons} G.\,W. Gibbons and S.\,W. Hawking,  
Phys. Rev. D {\bf 15}, 2738 (1977).
\bibitem{Anglin} J.\,R. Anglin and W. Ketterle, Nature {\bf 416}, 211 (2002).
\bibitem{SlowLight} W.\,G. Unruh and R. Sch\"utzhold, 
Phys. Rev. D {\bf 68}, 024008 (2003).
\bibitem{Garay} L.\,J. Garay, J.\,R. Anglin, J.\,I. Cirac, and 
P. Zoller, Phys. Rev. Lett. {\bf 85}, 4643 (2000).
\bibitem{CSM} C. Barcel\'o, S. Liberati, and M. Visser,
Class. Quantum Grav. {\bf 18}, 1137 (2001). 
\bibitem{Leonhardt} U. Leonhardt, T. Kiss, and P. \"Ohberg,
J. Opt. B 
{\bf 5}, S42  (2003).
\bibitem{Ourselves} P.\,O. Fedichev and U.\,R. Fischer, 
cond-mat/0303063.
\bibitem{BLV} C. Barcel\'o, S. Liberati, and M. Visser,
gr-qc/0305061. 
\bibitem{AQD} A. Recati {\it et al.}, 
cond-mat/0212413.
\bibitem{unruh76} W.\,G. Unruh, Phys. Rev. D {\bf 14}, 870 (1976).
\bibitem{BirrellDavies} N.\,D. Birrell and P.\,C.\,W. Davies, 
{\em Quantum Fields in Curved Space} (Cambridge University Press, Cambridge, 
England, 1984). 
\bibitem{fulling} S.\,A. Fulling,
Phys. Rev. D {\bf 7}, 2850 (1973).
\bibitem{unruh} W.\,G. Unruh, Phys. Rev. Lett. {\bf 46}, 1351 (1981).
\bibitem{Matt} M. Visser,  
Class. Quantum Grav. {\bf 15}, 1767 (1998).
\bibitem{PGRiemann} U.\,R. Fischer and M. Visser, 
Ann. Phys. (N.Y.) {\bf 304}, 22 (2003); 
Phys. Rev. Lett. {\bf 88,} 110201 (2002). 
\bibitem{Scaling} Yu. Kagan, E.\,L. Surkov,  and G.\,V. Shlyapnikov, 
Phys. Rev. A {\bf 54}, R1753 (1996); 
Y. Castin and R. Dum, Phys. Rev. Lett. {\bf 77}, 5315 (1996).
\bibitem{Zaremba} E. Zaremba, Phys. Rev. A {\bf 57}, 518 (1998).
\bibitem{No3} P.\,O. Fedichev and U.\,R. Fischer, 
cond-mat/0307200.
\bibitem{Stringari} S. Stringari, 
Phys. Rev. Lett. {\bf 77}, 2360 (1996). 
\bibitem{feshbach} A. Marte {\it et al.},  
Phys. Rev. Lett. {\bf 89}, 283202 (2002).
\end{thebibliography}
\end{document}